\documentclass[sigconf]{acmart}

\AtBeginDocument{%
  }

\usepackage{xcolor}

\copyrightyear{2026}
\acmYear{2026}
\setcopyright{cc}
\setcctype{by-nc-nd}
\acmConference[UMAP '26]{34th ACM Conference on User Modeling, Adaptation and Personalization}{June 08--11, 2026}{Gothenburg, Sweden}
\acmBooktitle{34th ACM Conference on User Modeling, Adaptation and Personalization (UMAP '26), June 08--11, 2026, Gothenburg, Sweden}
\acmDOI{10.1145/3774935.3806181}
\acmISBN{979-8-4007-2311-7/2026/06}

\usepackage{tabularx}
\usepackage{caption}
\usepackage{booktabs}
\usepackage{enumitem}
\usepackage{graphicx}
\usepackage{booktabs}       % 用于 toprule, midrule (美化表格线)
\usepackage{threeparttable} % 用于 tablenotes
\usepackage{enumitem}

\setlist[itemize]{topsep=2pt, partopsep=0pt, parsep=0pt}

\begin{document}

%%
%% The "title" command has an optional parameter,
%% allowing the author to define a "short title" to be used in page headers.
\title[Exploration of Diverse Movies]{How Personal Characteristics Shape User Exploration of Diverse Movie Recommendations with a LLM-Based Multi-Agent System}

\author{Yufan Zhou}
\authornote{Equal contribution.}
\email{yufan.zhou@kuleuven.be}
\affiliation{%
  \institution{Department of Computer Science, KU Leuven}
  \country{Belgium}}

\author{Yirui Huang}
\authornotemark[1]
\email{rae.yiruihuang@gmail.com}
\affiliation{%
  \institution{HII Lab, Duke Kunshan University}
  \country{China}}

\author{Zhao Wang}
\email{zhao\_wang@zju.edu.cn}
\affiliation{%
  \institution{Zhejiang University}
  \country{China}}

\author{Yucheng Jin}
\authornote{Corresponding author.}
\email{yucheng.jin@dukekunshan.edu.cn}
\affiliation{%
  \institution{HII Lab, Duke Kunshan University}
  \country{China}}

\renewcommand{\shortauthors}{Zhou et al.}

\begin{abstract}
  Diversity is an important evaluation criterion for recommender systems beyond accuracy, yet users differ in their willingness to engage with novel and diverse content. In this work, we investigate how a Large Language Model (LLM)–based multi-agent system supports users’ exploration of diverse recommendations, and how individual characteristics shape user experiences. We conducted a between-subjects user study (N = 100) comparing a multi-agent system with a single-agent system (baseline) for movie recommendations. We measured perceived accuracy, diversity, novelty, and overall rating, and examined the influence of personal characteristics, including personality traits, demographics, GenAI recommendation experience, and GenAI skepticism. Results show that the multi-agent system significantly increases perceived novelty and Shannon diversity. Conscientiousness is positively associated with perceived accuracy and diversity, whereas extraversion is negatively associated with perceived diversity. Prior experience with GenAI-based recommendations and age are positively associated with Shannon diversity, while skepticism toward GenAI is negatively associated with it. We also observe significant interaction effects between system design and user characteristics. These findings highlight the importance of personalization in multi-agent designs to support diverse recommendation exploration.
  % Results show that the multi-agent CRS significantly increases diversity measured by the Shannon index, compared to the single-agent baseline. Moreover, Agreeableness is positively associated with Perceived Accuracy, novelty, and recommendation rating, while Neuroticism (negatively) and prior experience with GPT (positively) are significantly associated with Shannon Diversity. These findings highlight the importance of personality-aware conversational recommender systems and caution against one-size-fits-all multi-agent designs.
  \end{abstract}




\maketitle

\section{Introduction}

In recommender systems research, diversity has long been regarded as an important evaluation dimension beyond accuracy~\cite{castells2021novelty, kaminskas2016diversity}: it not only determines whether users can access a broader content space, but also affects whether systems reinforce existing preferences and intensify “filter bubbles,” thereby limiting exploration and long-term satisfaction~\cite{nguyen2014exploring,areeb2023filter}. When recommender systems continuously focus on optimizing for users’ existing interests, they may improve click-through rates and perceived relevance in the short term, yet in the long run, such strategies often reduce users’ experiences of novelty, content coverage, and opportunities for self-expansion~\cite{liang2019recommender}.

A large body of research has proposed various approaches to increasing recommendation diversity from both algorithmic and user interface design perspectives~\cite{kunaver2017diversity,castells2021novelty,jin2018effects,hu2011enhancing,tsai2019exploring}. However, users may still be reluctant to explore diverse recommendations, even when such content is presented, due to limited attention and low interest in out-of-profile items~\cite{alves2024digitally}. The Large Language Models (LLMs) have demonstrated a great potential to proactively shape users’ exploration processes through explanations, comparisons, and interactive guidance~\cite{zhang2024see}.
Motivated by the potential of LLMs to support user exploration, we develop a multi-agent conversational recommender system (CRS) that simulates multiple movie-watching personas to nudge users toward diverse movie recommendations through personalized explanations. By enabling multiple agents to generate distinct rationales and viewpoints, the system can guide users to more openly weigh both in-profile and out-of-profile recommendations during the conversation. Unlike a single agent with a fixed narrative, multi-agent frameworks can produce a variety of recommendation explanations grounded in different personas and perspectives, which has been shown to enhance interaction experience and user preference modeling in recommender systems~\cite{fang2024multi}.

At the same time, an often overlooked fact is that users do not value diversity in the same way. Users differ in exploration motivation, risk tolerance, cognitive style, and decision strategies~\cite{sun2024interactive,liang2023promoting}, which leads to heterogeneous judgments about the value of diverse recommendations, different willingness to invest effort in out-of-profile content, and varying needs for explanation and persuasion. 
Prior work further suggests that personal characteristics (PC), such as personality traits and demographic factors, shape their preferences for novelty, controllability, and how information should be presented~\cite{jin2018effects,tang2022preference,kapoor2015like}. These findings suggest that diversity is not a uniform “add-on” that benefits all users equally, but rather an experiential goal that needs to be triggered, explained, and negotiated in a more personalized manner.
Accordingly, we examine whether the effectiveness of multi-agent interaction varies across users and how personal characteristics, including personality, demographics, prior experience with generative AI (GenAI) for recommendations, and skepticism about GenAI, may influence users’ experiences with diverse recommendation exploration.

We conducted a between-subjects online user study (N = 100) to compare a multi-agent conversational recommender system (CRS) with a single-agent baseline in terms of recommendation diversity and related metrics, including accuracy, novelty, and recommendation rating. We further examined how personal characteristics influence users’ experiences with the multi-agent system (MAS). The study addresses the following research questions:

\textbf{RQ1:} How does a multi-agent CRS influence diversity, novelty, accuracy, and recommendation rating?

\textbf{RQ2:} How do personal characteristics shape users’ experiences with a multi-agent CRS for exploring diverse recommendations?

The results show that the multi-agent CRS leads to significantly higher Perceived Novelty and Shannon Diversity compared to the single-agent baseline. In addition, Conscientiousness is positively associated with Perceived Accuracy and Perceived Diversity, whereas Extraversion is negatively associated with Perceived Diversity. Prior experience with GenAI-based recommendations and Age are positively associated with Shannon Diversity, while skepticism about GenAI is negatively associated with Shannon Diversity. Furthermore, we identify interaction effects between system design and selected personal characteristics on the Perceived Accuracy, Perceived Diversity, and Shannon Diversity.

This work makes three main contributions. \textit{First}, we design and implement an LLM-based multi-agent CRS to support users’ exploration of diverse movie recommendations. \textit{Second}, we provide empirical evidence demonstrating the effects of multi-agent design on recommendation diversity and related user-perceived metrics. \textit{Third}, we analyze how personal characteristics influence users’ experiences with the multi-agent CRS, highlighting the need for personalized multi-agent designs for exploring diverse recommendations.

\section{Related Work}

\subsection{LLMs for Conversational Recommendations}
%LLM-based CRS，现状：单一的
Recent advances in Large Language Models (LLMs) have fundamentally reshaped Conversational Recommender Systems (CRS). LLMs have evolved from serving as auxiliary components in discriminative recommendation pipelines to enabling end-to-end generative, multimodal, and agent-driven recommendation paradigms~\cite{shehmir2025llm4rec, liu2024large, hou2025survey}. LLM-driven recommender systems can perform flexible preference elicitation, adapt to evolving user intents, and generate fluent, personalized explanations, marking a clear departure from traditional slot-filling or rule-based conversational approaches ~\cite{mahmud2025evaluating, said2025explaining}.

Empirical and methodological work further demonstrates that LLM-based CRS improves conversational naturalness, supports implicit preference discovery, and enhances perceived recommendation quality~\cite{yun2025user}. However, existing evaluation frameworks and user studies largely assume a single conversational agent responsible for the entire interaction flow, reflecting a dominant single-agent design paradigm in current LLM-CRS research ~\cite{jannach2023evaluating, chen2025large}.

While prior work suggests that LLM-powered CRS can enhance user-system interaction by enabling more flexible and expressive interactions ~\cite{friedman2023leveraging}, such exploration is typically guided by a single reasoning pathway and a unified conversational persona. This single-agent assumption may limit perspective diversity and lead to homogeneous framing of recommendations, especially when users seek broader exploration or contrasting viewpoints~\cite{xia2025multi}. This limitation has motivated the design of multi-agent recommender systems powered by LLMs for exploring diverse recommendations.

\subsection{Approaches for Exploring Diverse Recommendations}
%可以回溯一些从交互帮助人们探索多元推荐的工作，应该有不少。可以从算法和界面的两个方面去写。
Diversity has always been a crucial goal beyond accuracy in recommendation systems. To help users explore more diverse content while maintaining relevance, existing work has generally followed two complementary angles: algorithmic diversification and interactive user interfaces.
% \subsubsection{Diversity-Enhancing Algorithms}

%（MMR）Multi‑armed RL；（DECREC）（MACRS）；LLM-based MAR
From an algorithmic perspective, classical diversification methods treat diversity as an item-level trade-off (e.g., MMR re-ranking~\cite{10.1145/290941.291025} and DPP-based ranking~\cite{kulesza2012determinantal}), while serendipity objectives further combine relevance with unexpectedness~\cite{10.1145/2926720}. Later work frames diversity as multi-objective optimisation and models exploration--exploitation via bandits and RL to optimise long-term utility~\cite{sun2018conversational,afsar2022reinforcement,wang2025diversity}. In conversational settings, multi-agent RL frameworks such as DECREC~\cite{wang2025diversity} and MACRS~\cite{fang2024multi} distribute conversational roles across specialised agents to sustain exploration over multiple turns. More recently, LLM-driven MAS orchestrate multiple recommendation/explanation personas~\cite{zhu2025llm, zhao2025exploring}, supporting diversity not only in items but also in viewpoints and explanations.

Interface-oriented work supports diverse exploration by exposing multiple facets of recommendations and giving users explicit control. Multi-list layouts present parallel carousels, which can increase Perceived Diversity and better serve heterogeneous goals, albeit sometimes at the cost of higher choice difficulty~\cite{tsai2017enhancing,tsai2019exploring,rahdari2022magic}. Relatedly, ``algorithmic affordances'' such as sliders, toggles, or buttons let users adjust objectives like diversity and exploration, improving transparency and user agency~\cite{smits2023results,jin2018effects_b,jin2020effects}.
In conversational recommenders, similar controls can be expressed through natural-language steering~\cite{kang2017understanding}. Recent evidence suggests that making diversity and novelty more controllable encourages users to engage in exploration and accept exploratory items~\cite{sun2024interactive}. 

While both algorithmic and interface-based approaches have demonstrated value in enhancing diversity, they often provide limited understanding of how such diversity-oriented designs are experienced by different users, particularly in conversational settings where explanations and social cues shape exploration. Crucially, most prior work operates at the \emph{retrieval} level, diversifying \textit{which} items are surfaced. A complementary yet underexplored question is whether users will actively engage with diverse candidates once they are presented. Our work targets this explanation-level gap: given an already-diversified candidate set, can multi-agent nudge users toward items outside their typical preferences?

\subsection{Impact of Personal Characteristics on User Experience with Recommender Systems}
%人格特质对用户感知交互的影响，chenli，chi22/cai，
Personality traits have emerged as significant moderators of user preferences in recommender systems, particularly for beyond-accuracy properties such as diversity. Prior work shows that personality captures stable user tendencies beyond interaction logs~\cite{ferwerda2016personality}, and that incorporating personality can improve perceived recommendation quality and satisfaction~\cite{nguyen2018user}.

Early studies further demonstrated that Big-Five traits are systematically associated with users’ needs for recommendation diversity in movie domains, suggesting that a single diversity level cannot fit all users~\cite{chen2013personality}. Follow-up work incorporated personality into diversification strategies, showing that personality-aware diversification better matches heterogeneous diversity preferences~\cite{10.1145/2481492.2481521}. Related studies further associate traits such as Openness and Extraversion with higher acceptance of diverse or complex content, whereas users lower in these traits tend to prefer more familiar recommendations~\cite{tintarev2013adapting,wu2018personalizing}. Moreover, users with different trait profiles may evaluate the same nominal diversity level differently, affecting satisfaction and perceived attractiveness~\cite{ferwerda2016influence}. Surveys highlight personality-aware recommendation as an ongoing area where diversity remains a central challenge and opportunity~\cite{dhelim2022survey}. 

%Beyond personality traits, prior research has shown that demographic factors such as age and gender also moderate how users perceive and interact with recommender systems, influencing click behavior and evaluation metrics~\cite{beel2013impact,neophytou2022revisiting,ekstrand2018all}, Perceived Diversity and novelty~\cite{wang2021user}. 
Beyond personality traits, recent studies on AI literacy show that users with higher AI knowledge and experience prefer more complex, detailed recommendation explanations and evaluate AI suggestions differently from less experienced users~\cite{ahokas2025influence}.
Broader research on GenAI further indicates that the duration and intensity of experience with GenAI tools systematically influence users’ trust calibration and how they interpret and act upon AI-generated outputs, including recommendations~\cite{djeric2025trust,sun2025revisiting}.

However, it remains unclear how personal characteristics, including personality, demographics, prior experience with GenAI, and skepticism about GenAI, could moderate users’ experience with multi-agent conversational recommenders for exploring diverse recommendations.

\section{Multi-Agent System Design}

\begin{figure*}[htbp]
    \centering
    \includegraphics[width=.9\textwidth]{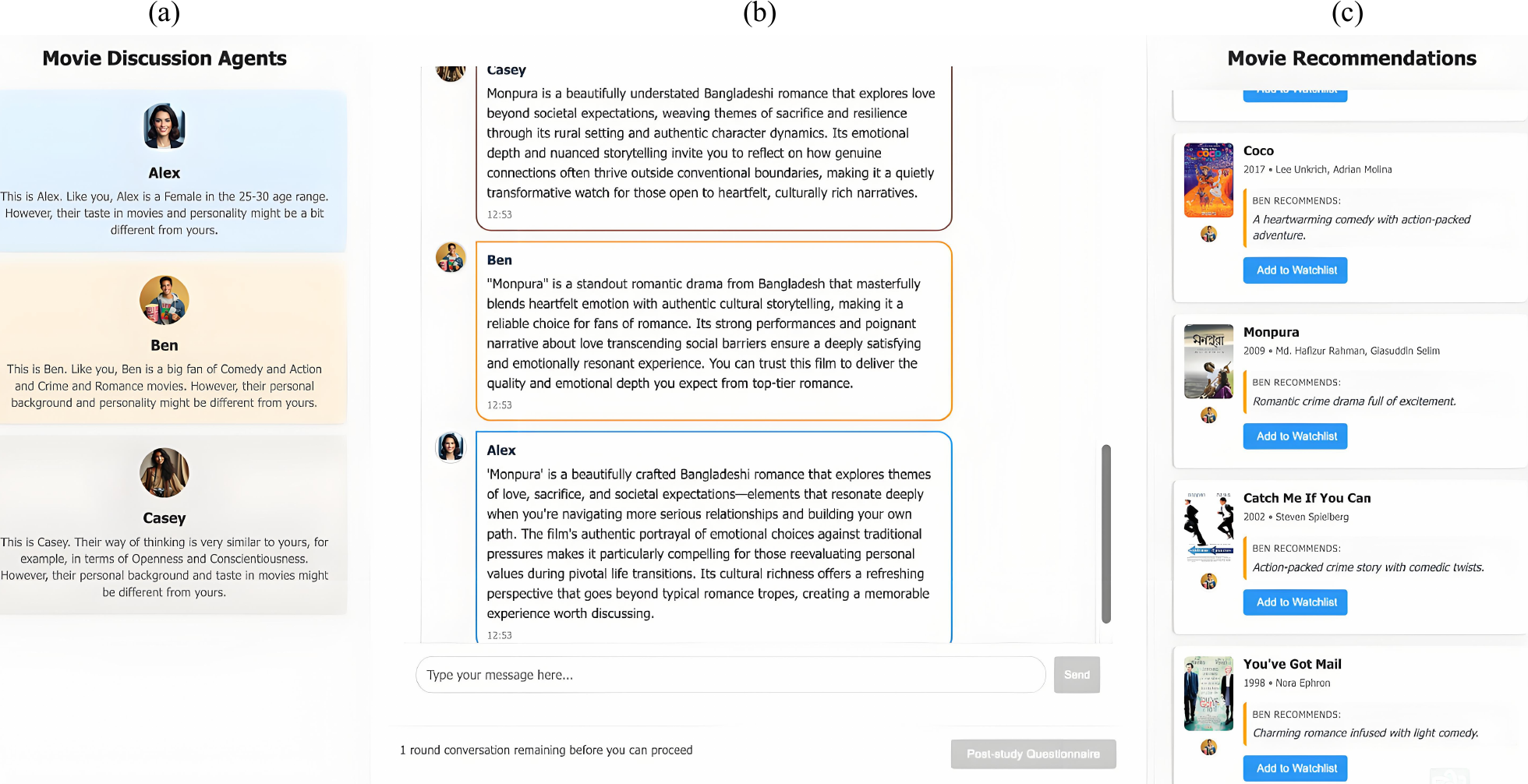}
    \caption{Main interface of the experimental system. The interface consists of three main panels. (a) \textbf{Agent Profile Panel:} Displays information about the conversational chatbot. (b) \textbf{Conversation Panel:} Shows the explanation dialogue between agents and the user. (c) \textbf{Movie Recommendation Panel:} The list of twelve recommended movies (6 in-profile, 6 off-profile), where users can view agent endorsements, access IMDb pages, add titles to their watchlist, and rate movies.}
    \label{fig:processes}
\end{figure*}

This section describes the design of a multi-agent system for generating diverse recommendations and simulating multiple personas that explain recommendations from different perspectives, potentially nudging users to explore items outside their typical preferences. We also describe the user interface design to illustrate how users interact with the system. All core code of this work has been open-sourced and is available at https://github.com/thzva/umap\_code.

\subsection{Diverse Recommendation Generation}
%To isolate the effect of explanation strategies on user exploration, we decoupled recommendation generation from explanation presentation. 

We utilized a curated corpus of 11,722 feature-length films (post-1970) derived from the IMDb non-commercial dataset\footnote{https://developer.imdb.com/non-commercial-datasets/} to generate diverse movie recommendations.
Based on a pre-study profile collected through a questionnaire administered before the main experiment (see Section~\ref{sec:procedure} for details), the system generates a personalized set of 12 movies for each participant using a controlled 6--6 split strategy. Specifically, the system randomly samples 2 preferred and 2 non-preferred genres from the participant's profile, then selects 3 films per genre to form 6 \textit{in-profile} and 6 \textit{off-profile} candidates. An LLM-based guard module verifies each candidate's contextual suitability against the participant's stated viewing scenario (e.g., time, occasion, companions), iteratively replacing unsuitable films until all candidates pass the check. The final 12 films are randomly reshuffled before presentation, ensuring a consistent candidate structure across participants while personalizing the specific film content to each individual. Crucially, this recommendation set remains fixed throughout the session. This design ensures that all users are exposed to a consistent level of item diversity, allowing us to investigate the impact of the explanation strategies (section~\ref{explanation_strategies}) on exploration of diverse recommendations.

\subsection{Persona Design for Explanations}
\label{explanation_strategies}
%We designed a multi-agent framework based on a \textit{Controlled Similarity} principle~\cite{stevens2016mimicry}.
Following prior work on the similarity–diversity trade-off in recommender systems~\cite{kaminskas2016diversity,zhou2010solving}, we balance relevance with diversity to support user exploration.
For each user, three distinct agents are constructed. Each agent matches the user on exactly one dimension---demographics (age and gender), movie preferences (genres), or personality---while deliberately differing on the other two. The recommendation list is generated before the explanation phase, and agents serve exclusively as \emph{explainers} that provide complementary and interpretive perspectives on the recommended items. Multiple agents may therefore positively discuss the same movie while emphasizing different reasons grounded in their persona-specific similarity with the user.
% The system assigns agents to explain a selected movie based on profile matching. For in-profile movies, explanations are provided by an agent who shares the user’s movie preference. For off-profile movies, explanations are provided by agents who do not share the same movie preference but align with the user along other dimensions, such as personality traits or demographic characteristics. 
% This assignment policy is designed to ensure that in-profile explanations reflect preference similarity, while off-profile explanations intentionally introduce alternative yet relatable perspectives. By doing so, the system supports users in exploring recommendations beyond their immediate preferences while maintaining a degree of personal relevance. Conversely, the single-agent baseline employs a fully matched persona aligning with the user across all three dimensions, representing an idealized "one-to-one" personalization scenario. 

To operationalize these personas, we employed a structured prompt engineering framework~\cite{robino2025conversation}. For each recommendation, parallel LLM calls generate explanations based on specific rhetorical guidelines: demographic agents focus on life-stage relevance, preference agents on genre alignment, and personality agents on cognitive or affective appeal. This approach ensures distinct, non-stereotypical reasoning for each persona. We present a representative excerpt from the prompt used for the demographic-matched agent (Agent A):
\begin{quote}
\footnotesize
\textbf{System Prompt (Agent A, Excerpt):}  
“You are Agent A (Alex), a movie enthusiast who shares demographic characteristics with the user. When generating explanations, you must focus on life-stage relevance inferred from the user’s profile, avoid explicit age references, and refrain from stereotypical assumptions. Your goal is to encourage exploration beyond the user’s comfort zone while maintaining an expert, analytical tone.”
\end{quote}

We validated the prompt design through an iterative qualitative evaluation before the main study. Two researchers independently assessed generated explanations for representative user profiles based on (1) diversity (distinct, persona-grounded reasoning across agents) and (2) meaningfulness (coherence and relevance). After each round, they discussed issues (e.g., generic or misaligned reasoning) and refined the prompts. This process continued until outputs showed consistent persona adherence and distinct, high-quality explanations, after which the prompts were fixed for the user study.

% We validated the effectiveness of the prompt design through an iterative qualitative evaluation process prior to the main user study. Two researchers independently reviewed explanation outputs generated for a set of representative user profiles and recommendation items. The outputs were evaluated along two criteria: (1) \emph{diversity of explanations}, i.e., whether different agents produced clearly distinguishable reasoning grounded in their assigned personas, and (2) \emph{meaningfulness}, i.e., whether the explanations were coherent, relevant, and informative for supporting user understanding.

% After each evaluation round, the researchers jointly discussed the outputs, identified failure cases (e.g., agents referencing unintended similarity dimensions, overly generic phrasing, or stereotypical reasoning), and refined the prompt instructions accordingly. This iterative process continued until the generated explanations consistently demonstrated stable persona adherence, non-overlapping explanatory strategies, and satisfactory explanatory quality. The finalized prompts were then fixed and used in the controlled user experiment.

\subsection{User Interface Design}
Figure~\ref{fig:processes} illustrates the web-based interface of the multi-agent conversational recommender system (CRS), which adopts a three-panel layout. The \textit{Movie Discussion Agents} panel displays three representative personas (Figure~\ref{fig:processes}[a]). Each persona card includes a portrait, name, and a description highlighting its similarities to and differences from the user.
The \textit{Conversation Panel} presents the explanations generated from a corresponding agent’s perspective (Figure~\ref{fig:processes}[b]). 
This design particularly allows users to see how others who share certain demographic or personality characteristics may endorse off-profile movies, thereby supporting exploration of more diverse recommendations.
Users can also ask follow-up questions to individual agents to further discuss a movie.
The \textit{Movie Recommendations} panel lists all movies that appear during the conversation. Each movie card displays metadata such as the title, director, actors, release year, and a link to the IMDb page for additional details. The card also indicates which agents endorse the movie, along with their brief opinions, and provides controls for users to add the movie to a watchlist and submit a star rating (Figure~\ref{fig:processes}[c]).
For the baseline condition, we keep the interface design consistent. The only difference is that the \textit{Movie Discussion Agents} panel displays a single agent, and the conversation panel presents only one agent’s explanation.

\section{User Experiment}
%This section introduces the experiment design, procedure, participants, and measures.
\subsection{Experiment Design}
We conducted a between-subjects study (N = 100) to examine how multi-agent design shapes users’ exploration experiences and perceptions of recommendations. Specifically, we compared a multi-agent system (MAS) with a single-agent baseline and investigated how the system design interacts with users’ personal characteristics to influence user experience outcomes, including diversity, accuracy, novelty, and recommendation rating.

To control for content effects, both conditions followed an identical recommendation structure. For each participant, a personalized set of 12 movies (6 in-profile and 6 off-profile) was generated at the start of the session based on their questionnaire responses and remained fixed throughout. No new items were introduced after this initial generation. The two conditions differed only in the number of agents and the explanation strategies used to support exploration. Below, we describe the experimental conditions.

\begin{itemize}[leftmargin=*]
    \item \textbf{C1: Single-Agent System (Baseline)} — A conversational recommender that provides recommendation-related information through a single agent whose profile closely aligns with the user’s personal characteristics. Explanations are delivered via a one-on-one conversational interaction.

    \item \textbf{C2: Multi-Agent System (MAS)} — A conversational recommender that includes multiple agents, each partially aligned with the user along one of three profile dimensions. These agents provide complementary perspectives on the recommended movies, enabling users to compare different viewpoints within the same conversational context.
\end{itemize}

\subsection{Experiment Procedure}
\label{sec:procedure}%回应审稿人的交叉引用要求
At the beginning of the study, participants read and provided informed consent (GDPR-compliant) and completed a pre-study questionnaire that collected demographic information, assessed personality traits, and asked participants to indicate 4–8 preferred movie genres. Participants were then randomly assigned to one of two conditions (C1 or C2; $N = 50$ per condition) and read a 90-second text-based tutorial introducing the system’s functionality and supported interactions. The task consisted of two phases:

\begin{itemize}[leftmargin=*]
\item \textbf{Phase 1 — Initial Recommendation.} Participants first described their viewing context (e.g., time, occasion, and companions). Based on this input, the system generated 12 movie recommendations (6 \emph{in-profile} and 6 \emph{out-of-profile}), accompanied by explanations according to the assigned condition (single-agent baseline in C1; MAS in C2).
\item \textbf{Phase 2 — Explanation-Only Dialogue.} After the recommendations were presented, participants could view the explanations generated by the corresponding agent(s) for each movie. Participants were able to ask follow-up questions about the recommended items, and the relevant agent responded to these queries.
\end{itemize}

To ensure sufficient engagement with the system, participants were required to complete at least five dialogue turns with the agents before proceeding to the rating phase. They then added movies of interest to their watchlist and rated at least three films on a 1--5 star scale. Participants were not required to rate all twelve movies. Participants then completed a post-study questionnaire. On average the whole session lasted approximately 20 minutes.
% \subsection{Experiment Design}
% 三个界面提供给用户进行交互，采取不同的交互策略。左侧界面展示了单代理解释过程，中间界面展示了多代理列表解释过程，右侧界面展示了多元代理社交解释过程。

% 单一中立代理在一对一聊天中提供文本电影推荐和解释。这作为比较的控制条件。
% 用户与三个基于配置文件的代理进行交互，这些代理显示为并排的卡片。代理人提供了不同的建议和理由：一个与个人资料内的选择相一致，而两个则使用人口统计/个性线索引入了个人资料外的选择。
% 这三个代理进行了实时、模拟的小组讨论，以决定推荐。轮流辩论是通过一个集成的提示和逐条消息的流式传输产生的，感觉就像一场实时聊天。

\subsection{Participants}
We recruited 100 participants via the online crowdsourcing platform Prolific.\footnote{https://www.prolific.com/} Participants were required to be fluent in English and located in the United States or the United Kingdom, and were compensated in accordance with Prolific’s fair wage policy. The study protocol received ethical approval from the Institutional Review Board (IRB) of the authors’ institution.

We included attention-check questions in both the pre- and post-study questionnaires to identify low-quality responses. In addition, we applied Prolific’s standard data-quality screening and replaced invalid or incomplete submissions until each condition (C1 and C2) reached 50 valid responses (N = 100).
The final sample was 57\% male and 43\% female. 
Participants were primarily in the 31--40 age group (30\%), with smaller proportions in 41--50 (26\%), 51--60 (19\%), > 60 (11\%), 26--30 (7\%), and 18--25 (7\%). Of the 100 participants, 57\% identified as male and 43\% as female. Regarding prior experience with GenAI-based recommendations, 49\% of participants reported never having used such systems, while 51\% had at least some experience, with the average level of usage at $M = 1.95$ ($SD = 1.06$) on a 5-point scale (1 = never, 5 = very frequently). Participants reported relatively low skepticism toward GenAI ($M = 1.52$, $SD = 1.57$ on a 0--5 scale, averaged across five items adapted from prior work).
% The age distribution skewed toward middle-aged and older adults, with 74\% of participants aged between 31 and 60. Overall, participants were active movie watchers, with over 79\% reporting that they watched movies at least once a week. Nearly 75\% reported using recommender systems ``sometimes'' or more frequently, and prior experience with generative AI recommendations was roughly balanced across the sample (50.7\% vs.\ 49.3\%).

\subsection{Measures}

\subsubsection{Personal Characteristics}

Before the main experiment, participants completed a pre-study questionnaire capturing personal characteristics, including (1) gender and age range; (2) movie preferences (selecting 4--8 of 21 genres) and viewing frequency; and (3) general recommender usage, and prior use of generative AI (e.g., ChatGPT, Claude, Gemini) for recommendations. Personality was measured using the Ten-Item Personality Inventory for the Big Five (Openness, Conscientiousness, Extraversion, Agreeableness, Neuroticism) on a 5-point Likert scale~\cite{rammstedt2007measuring}. 
%Table~\ref{tab:personality_descriptive} shows the detailed results about participants' personality. 

\subsubsection{Recommendation Diversity}

\begin{table}[H]
\centering
\footnotesize % 保持小字号
\renewcommand{\arraystretch}{1.1} % 1.1 倍行距，紧凑但即使只有单倍行距也不拥挤
\setlength{\tabcolsep}{3pt} % 稍微收紧列宽

\caption{Standardized Loadings, Reliability, and Convergent Validity for the Metrics }
\label{tab:questionnaire_quality_compact}

% 改为两列布局：X (描述) 和 c (Loading)
% 这样 X 列更宽，长句子更不容易换行，能节省高度
\begin{tabularx}{\columnwidth}{>{\raggedright\arraybackslash}X c}
\toprule
\textbf{Construct / Item Description} & \textbf{Loadings} \\
\midrule

% === Construct 1 ===
% 将标题和数据合并在一行，并用小号字显示数据，争取一行放下
\textbf{Perceived Accuracy} {\footnotesize ($\alpha$=0.926; CR=0.954; AVE=0.874)} & \\
\hspace{1em} The recommended movies were well-chosen. & 0.963 \\
\hspace{1em} The recommended movies seemed relevant to my situation. & 0.935 \\
\hspace{1em} The recommended movies were interesting. & 0.906 \\
\addlinespace[0.4em] % 稍微一点点间距区分模块，不要太大

% === Construct 2 ===
\textbf{Perceived Novelty} {\footnotesize ($\alpha$=0.889; CR=0.932; AVE=0.821)} & \\
\hspace{1em} The chatbot helped me discover new movies. & 0.889 \\
\hspace{1em} The chatbot provided me with surprising recommendations. & 0.916 \\
\hspace{1em} The chatbot provided recommendations that were a pleasant surprise. & 0.913 \\
\addlinespace[0.4em]

% === Construct 3 ===
\textbf{Perceived Diversity} {\footnotesize ($\alpha$=0.879; CR=0.925; AVE=0.805)} & \\
\hspace{1em} The chatbot has a more varied selection of movies. & 0.895 \\
\hspace{1em} The recommended movies match a wide variety of moods. & 0.909 \\
\hspace{1em} The recommendations would suit a broad set of tastes. & 0.888 \\

\bottomrule
\end{tabularx}
\end{table}

We assessed recommendation diversity using both subjective and behavioral measures. Subjective diversity was measured using three validated items from the \textit{CRS-Que} evaluation framework (see Table~\ref{tab:questionnaire_quality_compact})~\cite{jin2024crs}, which capture users’ perceptions of whether the system offers a varied set of recommendations that span different moods and tastes and avoid repetitive suggestions.  

In addition, we computed Shannon Diversity (H)~\cite{ziegler2005improving} to quantify diversity based on users’ selection behavior. This metric captures the richness and evenness of genre distribution within the aggregated set of movies selected by users. Higher values indicate that users, as a group, selected items from a broader and more balanced range of genres. The metric is defined as:
\[
H = -\sum_{i=1}^{S} p_i \ln(p_i),
\]
where $S$ denotes the number of unique genres in the selected items, and $p_i$ represents the proportion of items belonging to genre $i$. We also computed related diversity metrics, including Simpson Diversity and Intra-List Diversity (ILD), which showed consistent trends, supporting the robustness of our findings. This dual measurement strategy allows us to disentangle whether the multi-agent framework influences users' actual selection behavior versus their subjective perception of diversity.

To further validate the aggregate-level Shannon metric, we conducted a supplementary participant-level diversity analysis, computing Shannon entropy, Simpson index, ILD, and rating variability for each individual's selections. Mann-Whitney U tests revealed no significant between-condition differences on any individual-level metric (all $p > .05$). Moreover, individual-level metrics showed only weak correlations with the aggregate Shannon index ($|r| \leq 0.15$, all $p > .05$), suggesting that these two levels of analysis capture distinct facets of diversity: the aggregate metric reflects recommendation-list-level genre coverage, whereas individual-level metrics capture personal selection patterns. The full analysis scripts and results are available in the code repository.

\subsubsection{Other Diversity-Related Measures}
We further employed validated items from \textit{CRS-Que} (Table~\ref{tab:questionnaire_quality_compact}) to measure \textit{Perceived Novelty} and \textit{Perceived Accuracy}, two constructs commonly associated with diversity in recommender systems. \textit{Perceived Novelty} captures the extent to which recommendations are surprising and help users discover new content, while \textit{Perceived Accuracy} reflects how well the recommended movies align with users’ preferences and viewing context. Finally, we assessed overall recommendation quality using users’ ratings of the movies they added to their watchlists.

\section{Results and Analysis}
We employed linear regression models with system condition as the independent variable and users’ personal characteristics as covariates, and five diversity-related metrics as dependent variables. 

% We first examine whether, and to what extent, the multi-agent setting (Condition C2) differs from the single-agent baseline (Condition C1) across the full set of evaluation metrics. In the regression models, Condition C1 is treated as the reference category, so the estimated coefficient for Condition C2 captures the expected change in each metric attributable to the multi-agent design, holding the remaining covariates constant. The specific results are shown in the Table ~\ref{tab:regression_results}
\subsection{Effects of Multi-Agent Design}

\begin{table*}[htbp]
\centering
\scriptsize
\begin{threeparttable}
\caption{Regression Models: Effects of Condition, Personality, and Covariates on User Perceptions}
\label{tab:regression_main}
\begin{tabular}{lccccc}
\hline
 & \multicolumn{3}{c}{\textbf{Subjective Metrics}} & \multicolumn{2}{c}{\textbf{Objective Metrics}} \\
\cmidrule(lr){2-4} \cmidrule(lr){5-6}
 & \textbf{Accuracy} & \textbf{Novelty} & \textbf{Diversity} & \textbf{Shannon} & \textbf{Avg.~Rating} \\
 & Coef. (S.E.) & Coef. (S.E.) & Coef. (S.E.) & Coef. (S.E.) & Coef. (S.E.) \\
\hline
Condition (C1 vs.\ C2) & 0.643 (0.651) & \textbf{1.507 (0.720)}* & 0.003 (0.644) & \textbf{0.546 (0.202)}** & 0.230 (0.564) \\
\hline
Openness & $-$0.013 (0.136) & $-$0.206 (0.150) & $-$0.108 (0.134) & $-$0.026 (0.042) & $-$0.014 (0.117) \\
Conscientiousness & \textbf{0.526 (0.209)}* & 0.008 (0.231) & \textbf{0.462 (0.207)}* & 0.102 (0.065) & 0.281 (0.181) \\
Extraversion & $-$0.085 (0.149) & 0.013 (0.165) & \textbf{$-$0.307 (0.147)}* & 0.060 (0.046) & $-$0.124 (0.129) \\
Agreeableness & $-$0.226 (0.219) & 0.066 (0.242) & $-$0.020 (0.216) & $-$0.022 (0.068) & $-$0.110 (0.189) \\
Neuroticism & 0.098 (0.164) & 0.176 (0.181) & 0.055 (0.162) & 0.001 (0.051) & 0.129 (0.142) \\
\hline
Condition $\times$ Openness & $-$0.157 (0.226) & 0.086 (0.250) & $-$0.153 (0.224) & 0.045 (0.070) & 0.010 (0.196) \\
Condition $\times$ Conscientiousness & \textbf{$-$0.574 (0.261)}* & $-$0.054 (0.288) & $-$0.463 (0.258)$^{\dagger}$ & $-$0.122 (0.081) & $-$0.235 (0.226) \\
Condition $\times$ Extraversion & 0.205 (0.202) & 0.178 (0.223) & \textbf{0.406 (0.200)}* & $-$0.093 (0.063) & 0.092 (0.175) \\
Condition $\times$ Agreeableness & \textbf{0.674 (0.258)}* & 0.254 (0.286) & 0.207 (0.256) & 0.024 (0.080) & 0.365 (0.224) \\
Condition $\times$ Neuroticism & 0.043 (0.233) & $-$0.095 (0.258) & 0.088 (0.230) & 0.027 (0.072) & $-$0.029 (0.202) \\
\hline
GenAI Skepticism & 0.030 (0.183) & $-$0.087 (0.203) & 0.132 (0.182) & \textbf{$-$0.120 (0.057)}* & 0.092 (0.159) \\
GenAI Recommendation Experience & 0.337 (0.262) & 0.474 (0.290) & 0.056 (0.259) & \textbf{0.275 (0.081)}** & $-$0.013 (0.227) \\
Age & 0.068 (0.112) & 0.209 (0.124)$^{\dagger}$ & $-$0.095 (0.111) & \textbf{0.082 (0.035)}* & $-$0.013 (0.097) \\
Gender & 0.334 (0.295) & $-$0.120 (0.327) & $-$0.068 (0.292) & $-$0.053 (0.092) & $-$0.022 (0.256) \\
\hline
Condition $\times$ GenAI Skepticism & 0.222 (0.236) & 0.253 (0.261) & $-$0.052 (0.233) & 0.135 (0.073)$^{\dagger}$ & $-$0.106 (0.204) \\
Condition $\times$ GenAI Recommendation Experience & $-$0.575 (0.346) & $-$0.589 (0.382) & 0.005 (0.342) & \textbf{$-$0.250 (0.107)}* & 0.160 (0.300) \\
Condition $\times$ Age & $-$0.072 (0.154) & $-$0.278 (0.171) & $-$0.009 (0.153) & \textbf{$-$0.101 (0.048)}* & $-$0.090 (0.134) \\
Condition $\times$ Gender & $-$0.422 (0.405) & $-$0.046 (0.448) & 0.235 (0.401) & $-$0.018 (0.126) & 0.015 (0.351) \\
\hline
Constant & 3.436 (0.443)*** & 2.981 (0.489)*** & 4.310 (0.438)*** & 0.480 (0.137)*** & 4.085 (0.383)*** \\
$R^2$ & 0.311 & 0.230 & 0.242 & 0.307 & 0.173 \\
Adjusted $R^2$ & 0.147 & 0.047 & 0.062 & 0.142 & $-$0.023 \\
\hline
\end{tabular}
\begin{tablenotes}\scriptsize
\item \textit{Note.} $N = 100$. Condition C2 (multi-agent) is compared to Condition C1 (single-agent).
Personality traits and AI-related covariates are mean-centered.
Age is coded as an ordinal category (1 = 18--25, 2 = 26--30, 3 = 31--40, 4 = 41--50, 5 = 51--60, 6 = over 60).
Gender is coded as 1 = male, 0 = female.
Significant coefficients ($p < .05$) are shown in boldface.
\item Significance levels: *** $p < .001$, ** $p < .01$, * $p < .05$, $^{\dagger}$ $p < .10$.
\end{tablenotes}
\end{threeparttable}
\end{table*}

% \begin{table}[htbp]
%   \footnotesize
%   \centering
%   \caption{Descriptive Statistics of Dependent Variables}
%   \label{tab:dv_descriptive}
%   \small
%   \begin{tabular}{lccc}
%     \toprule
%     {\small Measure} & {\small Cond. A} & {\small Cond. B} & {\small Overall} \\
%     {\small } & {\small M (SD)} & {\small M (SD)} & {\small M (SD)} \\
%     \midrule
%     Accuracy & 3.83 (1.02) & 3.96 (0.99) & 3.90 (0.99) \\
%     Novelty & 3.72 (1.05) & 4.09 (1.01) & 3.91 (1.04) \\
%     Diversity & 3.88 (0.95) & 3.97 (0.92) & 3.91 (0.94) \\
%     Shannon & 0.76 (0.30) & 0.92 (0.30) & 0.84 (0.31) \\
%     Avg Rating & 3.97 (0.70) & 3.88 (0.86) & 3.93 (0.79) \\
%     \bottomrule
%   \end{tabular}
%   \begin{tablenotes}
% \scriptsize
% \item \textit{Note.} M = Mean; SD = Standard Deviation; Cond. = Condition.
%   \end{tablenotes}
% \end{table}

% \vspace{-1.5em}

\begin{table}[H]
  \footnotesize % 统一设置表格字体大小
  \centering
  \begin{threeparttable}
  \caption{Descriptive Statistics of Dependent Variables}
  \label{tab:dv_descriptive}
  \begin{tabularx}{\linewidth}{lXXX} % 使用 tabularx 并设置三列等宽自适应
    \toprule
    Measure & Cond. C1 & Cond. C2 & Overall \\
            & M (SD)  & M (SD)  & M (SD)  \\
    \midrule
    Accuracy   & 3.83 (1.02) & 3.96 (0.99) & 3.90 (0.99) \\
    Novelty    & 3.72 (1.05) & 4.09 (1.01) & 3.91 (1.04) \\
    Diversity  & 3.88 (0.95) & 3.97 (0.92) & 3.91 (0.94) \\
    Shannon    & 0.76 (0.30) & 0.92 (0.30) & 0.84 (0.31) \\
    Avg Rating & 3.97 (0.70) & 3.88 (0.86) & 3.93 (0.79) \\
    \bottomrule
  \end{tabularx}
  \begin{tablenotes}
    \scriptsize
    \item \textit{Note.} M = Mean; SD = Standard Deviation; Cond. = Condition.
  \end{tablenotes}
  \end{threeparttable} 
\end{table}

% \vspace{-1.5em}

Table~\ref{tab:regression_main} reports the regression results, illustrating how diversity-related user perceptions are associated with system design and personal characteristics, as well as their interaction effects, represented by interaction terms in the models. We report regression coefficients, standard errors, $p$-values, and both $R^2$ and adjusted $R^2$ values. 

%We examined the main effects of the multi-agent design relative to the single-agent baseline across five dependent variables (Table~\ref{tab:regression_main}). 

% Specifically, the regression results revealed a significant positive effect of the multi-agent condition on Shannon Diversity ($\beta = 0.546, SE = 0.202, p < .01$), indicating that users in the multi-agent condition selected movies from a more diverse set of genres. In contrast, we did not observe a significant main effect on Perceived Diversity ($\beta = 0.003, SE = 0.644, p > .05$).

% The multi-agent design also significantly increased Perceived Novelty ($\beta = 0.437, p < .05$). these results suggest that while users’ explicit perceptions of “diversity” did not significantly change, the multi-agent design was associated with more behaviorally diverse selections and higher Perceived Novelty.删除

% The multi-agent design significantly increased Perceived Novelty ($\beta = 1.507, SE = 0.720, p < .05$), indicating that users in the multi-agent condition found the recommendations more surprising and discovery-oriented. Moreover, the multi-agent condition led to significantly higher Shannon Diversity ($\beta = 0.546, SE = 0.202, p < .01$), reflecting a broader and more balanced genre distribution in users' selections.}

% However, the multi-agent condition did not significantly affect Perceived Diversity ($\beta = 0.003, SE = 0.644, p > .05$), suggesting that while objective diversity and novelty increased, users' explicit judgments of ``diversity'' remained unchanged.

Overall, the multi-agent condition (C2) showed higher scores than the baseline condition (C1) on most measures (see Table~\ref{tab:dv_descriptive}).
Specifically, the multi-agent design significantly increased Perceived Novelty ($\beta = 1.507, SE = 0.720, p < .05$), indicating that users in the multi-agent condition found the recommendations more surprising and discovery-oriented. The multi-agent condition also led to significantly higher Shannon Diversity ($\beta = 0.546, SE = 0.202, p < .01$), reflecting a broader and more balanced genre distribution in users' selections. However, the multi-agent condition did not significantly affect Perceived Diversity ($\beta = 0.003, SE = 0.644, p > .05$).
%, suggesting that while objective diversity and novelty increased, users' explicit judgments of ``diversity'' remained unchanged.
Furthermore, the multi-agent condition did not yield significant main effects on Perceived Accuracy ($\beta = 0.643, SE = 0.651, p > .05$) or Average Rating ($\beta = 0.230, SE = 0.564, p > .05$). This indicates that the observed increases in novelty and objective diversity were not accompanied by systematic decreases in Perceived Accuracy or overall user rating.

\subsection{Effects of Personal Characteristics}

\begin{figure*}[htbp]
    \centering
    \includegraphics[width=\textwidth]{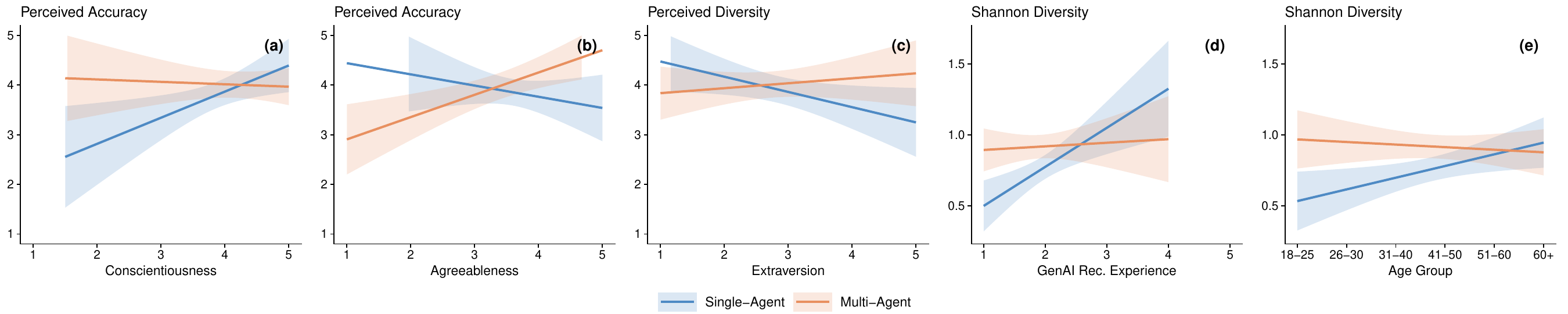}
    \caption{Interaction Effects of Condition and Personality Traits on User Perceptions}
    \label{fig:interaction_effects}
\end{figure*}
%先report pc 对于不同measure直接的相关性
Next, we report the direct correlation between users' personal characteristics and the measures for diversity, accuracy, novelty, and recommendation rating.

\textbf{Personality Traits.} 
\textit{Conscientiousness} was positively associated with both Perceived Accuracy ($\beta = 0.526, SE = 0.209, p < .05$) and Perceived Diversity ($\beta = 0.462, SE = 0.207, p < .05$).
In contrast, \textit{Extraversion} was negatively associated with Perceived Diversity ($\beta = -0.307, SE = 0.147, p < .05$), indicating that more extraverted users may have a higher threshold for perceiving content as diverse. Other personality traits, including \textit{Openness}, \textit{Agreeableness}, and \textit{Neuroticism}, did not exhibit significant main effects across the evaluated outcomes. 
% A marginally significant positive association was also observed between \textit{Conscientiousness} and Shannon Diversity ($\beta = 0.102, SE = 0.065, p < .1$), suggesting a potential tendency for more conscientious users to select from a broader range of genres.

\textit{GenAI Recommendation Experience} was positively associated with objective Shannon Diversity ($\beta = 0.275, SE = 0.081, p < .01$), indicating that users with more experience in GenAI potentially tend to follow agents' explanations to explore off-profile recommended movies. In contrast, \textit{GenAI Skepticism} was negatively associated with Shannon Diversity ($\beta = -0.120, SE = 0.057, p < .05$), suggesting that users with greater skepticism toward GenAI selected from a narrower range of genres. \textit{Age} was also positively associated with Shannon Diversity ($\beta = 0.082, SE = 0.035, p < .05$). \textit{Gender} did not yield significant main effects.
%and \textit{Gender} did not yield significant main effects.

\subsection{Interaction Effects}

%描述pc和condition的所有交互效应
Finally, we report how personal characteristics interact with system design conditions to influence dependent variable measures. The results indicate heterogeneity in how users respond to the multi-agent mechanism, with notable moderation effects on Perceived Accuracy and marginal trends on diversity and average rating.

\textbf{Moderation by Conscientiousness.} 
Conscientiousness moderated the relationship between Condition and Perceived Accuracy ($\beta = -0.574, SE = 0.261, p < .05$), indicating that the strength/direction of the association differed across levels of Conscientiousness (Figure~\ref{fig:interaction_effects} [a]).
Although conscientious users generally rated the system more positively regarding accuracy (main effect), the relative advantage of the multi-agent condition (vs. single-agent) decreased as Conscientiousness increased, and may even reverse at higher levels of Conscientiousness. A similar but marginal pattern was found for Perceived Diversity ($\beta = -0.463, SE = 0.258, p < .1$), suggesting a potential mismatch between the multi-agent exploration strategy and the expectations of highly conscientious users.

\textbf{Moderation by Agreeableness.}
We found a significant interaction between Agreeableness and Condition on Perceived Accuracy ($\beta = 0.674, SE = 0.258, p < .05$). As illustrated in Figure~\ref{fig:interaction_effects} (b), users with higher Agreeableness perceived the MAS as more accurate than the single-agent baseline, whereas the estimated benefit was smaller among lower-Agreeableness users.
%The interaction on Average Rating was in the same direction but did not reach significance ($\beta = 0.365, SE = 0.224, p > .1$).}

\textbf{Moderation by Extraversion.} 
Finally, we observed a significant positive interaction between Extraversion and Condition on Perceived Diversity ($\beta = 0.406, SE = 0.200, p < .05$). This indicates that the multi-agent design mitigated the lower diversity perceptions typically associated with more extraverted users, making the system's exploratory behavior more acceptable to this group. The visualization results are shown in Fig.~\ref{fig:interaction_effects} (c).

\textbf{Moderation by GenAI-Related Attitudes and Age.} 
We observed several interactions between Condition and user covariates on Shannon Diversity. The interaction between Condition and \textit{GenAI Recommendation Experience} was significantly negative ($\beta = -0.250, SE = 0.107, p < .05$), indicating that the diversity advantage associated with prior GenAI experience was attenuated in the multi-agent condition. The interaction between Condition and \textit{Age} was also significantly negative ($\beta = -0.101, SE = 0.048, p < .05$), suggesting that the diversity benefit of age was reduced under the multi-agent design. In addition, the interaction between Condition and \textit{GenAI Skepticism} was marginally positive ($\beta = 0.135, SE = 0.073, p < .1$), suggesting that the multi-agent design may have partially offset the negative association between skepticism and genre diversity.

\section{Discussion}
This study investigates an LLM-based multi-agent system (MAS) in movie recommendation, examining (1) the overall effects of multi-agent versus single-agent explanations (RQ1) and (2) how personal characteristics, particularly personality traits, shape users’ experience of exploring diversity (RQ2).
Results show that multi-agent design supports broader exploration, increasing \emph{Perceived Novelty} and objective diversity, though not users’ perceived diversity. These effects are further moderated by personal characteristics, including personality traits (e.g., Conscientiousness, Extraversion, Agreeableness), prior GenAI experience, and age.

%1.解读一下multi-agent设计对于diversity的作用，例如主管主观measure没有显著，客观香农measure有用意味着什么。novelty提升意味着什么？accuracy和rating没有下降，diversity的nudge没有对推荐质量造成负面的现象？可以找一下文献进一步解释我们的结果；以及和之前support recommendation diversity的研究工作结果做一些对比，看看有什么不同，我们有没有什么新的发现。尤其这个是在llm driven下面产生的

%2。同上解释personal characteristics的结果，也是找一些可能的文献去解释和支撑。如果结论不一样的话可以写一下argument，讲讲实验直接的不同设置可能会造成不一致。另外就是对于交互效应的解释，可以提出一些design implication，针对某一类人，multi-agent design需要怎么适配

\subsection{Accuracy--Diversity Trade-off}

A longstanding concern in recommender design is the \textit{accuracy--diversity trade-off}, where pushing beyond-accuracy objectives may risk degrading perceived relevance or overall user evaluations~\cite{kaminskas2016diversity}. In our study, the MAS significantly increased both objective diversity (Shannon Diversity) and Perceived Novelty, while showing no significant main-effect differences in Perceived Accuracy or overall user ratings. Rather than demonstrating that the trade-off disappears, these results indicate that \emph{explanation-level} multi-agent reasoning can broaden the recommendation distribution without a systematic penalty in global quality perceptions---at least for the movie exploration task studied here.

One plausible account is that the multi-agent interaction operates differently from classic diversification methods (e.g., re-ranking with diversity constraints)\cite{carraro2025enhancing}. Instead of explicitly penalizing similar or high-confidence items, the agents generate and negotiate alternative interpretations of the user's preferences in natural language, which can surface adjacent options and expand the item exposure distribution as a \emph{byproduct} of reasoning. This aligns with emerging evidence that LLM-powered CRS increasingly shape outcomes through interaction and language, not only through scoring or ranking pipelines~\cite{jannach2021survey, guo2023towards}.
Notably, we observed a significant increase in Shannon Diversity, although no significant main effect was found for Perceived Diversity.
This dissociation echoes prior work showing that objective diversity metrics do not always translate into user-Perceived Diversity~\cite{ekstrand2014user}, which is often driven by salient, ``intuitive'' variety cues (e.g., distinct genres) and can saturate beyond a threshold~\cite{dokoupil2024user}. In this light, our results suggest that multi-agent interaction may primarily change the \emph{structure} of exposure (entropy) and the \emph{felt experience of surprise} (novelty), while leaving the explicit label of "diversity" unchanged unless the interface makes category contrasts more legible~\cite{hu2011helping}. This suggests that for LLM-based CRS, multi-agent reasoning alone may not make diversity perceptible; it likely requires complementary \emph{presentation} cues (e.g., grouping, contrast, or facet highlighting)~\cite{jesse2023intra, hu2011helping}.

\subsection{Personality for Multi-Agent Interaction}
The interaction effects reveal that the ``one-size-fits-all'' approach could be risky in multi-agent interface design. These findings regarding personal characteristics echo prior work showing that users differ in their needs for diversity in movie recommendations~\cite{sun2024interactive}.

%The same debate mechanism that delighted some users alienated others~\cite{alves2022examining, alves2020incorporating}.

\textbf{Conscientiousness and the Need for Efficiency.}
Users with high Conscientiousness penalized the MAS, perceiving it as significantly less accurate. The literature on personality computing suggests that conscientious individuals prioritize efficiency, order, and task completion~\cite{john2010handbook}. The multi-agent design, which exposes users to conflicting viewpoints and a discussion process, may introduce cognitive load and perceived uncertainty~\cite{chen2019managing}. For a user seeking a ``correct answer'' efficiently, the discussion might be viewed as noise or indecisiveness rather than value-added elaboration. Prior work shows that conscientious users prefer transparent, deterministic interfaces over complex, exploratory ones~\cite{naiseh2023different}.

\textbf{Agreeableness and Social Trust.}
In contrast, highly Agreeable users perceived the MAS as significantly more accurate. Agreeable individuals are characterized by a tendency towards social harmony and trust~\cite{john2010handbook}. We argue that these users tend to interpret the multi-agent interaction as a cooperative social effort to serve them better. They may appreciate the ``social presence'' of multiple agents working together, viewing the consensus reached by agents as a stronger signal of quality than a single output. This aligns with the \textit{Media Equation} theory~\cite{sharan2020effects}, suggesting that agreeable users apply social rules of politeness and trust to the collaborative agents.

\textbf{Extraversion and the Need for Stimulation.}
Our results also indicated that Extraverts, who typically have a higher threshold for stimulation, benefited from the multi-agent condition in terms of Perceived Diversity~\cite{philipp1970stimulation,eysenck1991dimensions}. The baseline condition (C1) might have been too static or predictable for them. The dynamic interplay of agents provided the necessary variance and novelty to satisfy their intrinsic need for exploration.

\subsection{Implications for Adaptive Design}

These findings suggest that multi-agent recommender systems should move beyond uniform interaction strategies and instead adopt \textit{personalized and adaptive designs}. 
For highly conscientious users, interfaces should minimize the visibility of internal deliberation processes and emphasize concise, consensus-driven outputs. Presenting a clear recommendation with a compact rationale may better align with their preference for efficiency and certainty.
For agreeable (and potentially open) users, exposing the collaborative dynamics among agents may enhance trust and engagement. Making the interaction process transparent can help transform the system from a black box into an interpretable and trustworthy partner.
Finally, the positive association between user experience with GenAI and objective diversity suggests that experienced users learn to actively steer agents toward different regions of the item space. To support novice users, future systems could provide scaffolded prompt guidance~\cite{zhang2026cross}, such as predefined exploration trajectories, enabling broader access to the exploratory potential of MAS.

\section{Limitations}
This work has three key limitations.
\textit{First}, our personality-aware MAS relies on a static user profile derived from pre-study questionnaires (TIPI), which minimizes participant burden but captures only coarse trait signals. Future work could leverage conversational interactions to dynamically infer and update user traits, enabling more adaptive, interaction-driven personalization.
\textit{Second}, to isolate the effects of multi-agent explanation strategies, we decoupled retrieval from explanation by presenting a fixed set of twelve movies. While this ensures internal validity, it limits ecological realism, as agent interactions in practice would likely influence retrieval itself. Nevertheless, our core mechanism—using complementary personas and contrasting rationales to promote exploration—can be integrated with dynamic recommendation pipelines, which we leave for future work.
\textit{Finally}, generalizability is limited by our sample (US/UK Prolific users) and single-session design. Cultural factors may influence responses to explanations, and it remains unclear whether observed shifts toward off-profile exploration persist over time, warranting longitudinal investigation.

\section{Conclusion}
This study examined how multi-agent conversational recommender design and user characteristics jointly shape user perceptions. The results demonstrate that the effectiveness of system design is highly dependent on users' personality traits, and that ``one-size-fits-all'' designs are insufficient to satisfy diverse user needs.
Specifically, the multi-agent design improved both perceived novelty and objective diversity, but its effects were not uniform across users. Instead, outcomes were moderated by personal characteristics, with certain traits amplifying perceived benefits while others attenuated them.
These findings suggest that future LLM-based recommenders should move beyond content personalization toward interaction-level adaptation, tailoring explanation and collaboration strategies to individual traits. Personality-aware adaptive design offers a promising pathway to encourage more diverse exploration and to mitigate the filter-bubble effects commonly associated with recommender systems.

\begin{acks}
We used AI-assisted tools (e.g., GPT, Gemini, and Claude) to assist with manuscript writing and system development. All methodological decisions, analyses, and conclusions are solely those of the authors. This work was supported by the startup research grant of Duke Kunshan University (00AKUG0250), the Kunshan Shuangchuang Research Grant (KSSC202501020), and the Key R\&D Program of Ningbo (2024Z254).
\end{acks}

\bibliographystyle{ACM-Reference-Format}
\bibliography{sample-base}

\end{document}